# Low dose X-ray speckle visibility spectroscopy reveals nanoscale dynamics in radiation sensitive ionic liquids

Jan Verwohlt[1], Mario Reiser[1,2], Lisa Randolph[1], Aleksandar Matic[3], Luis Aguilera Medina[3], Anders Madsen[2], Michael Sprung[4], Alexey Zozulya[2,4] and Christian Gutt[1*]

[1] Department Physik, Universität Siegen, D-57072 Siegen, Germany
[2] European X-Ray Free-Electron Laser Facility, D-22869 Schenefeld, Germany
[3] Department of Physics, Chalmers University of Technology, 41296 Gothenburg, Sweden
[4] Deutsches Elektronen-Synchrotron DESY, Hamburg, Germany

[*] gutt@physik.uni-siegen.de

**Abstract**

*X-ray radiation damage provides a serious bottle neck for investigating µs to s dynamics on nanometer length scales employing X-ray photon correlation spectroscopy. This limitation hinders the investigation of real time dynamics in most soft matter and biological materials which can tolerate only X-ray doses of kGy and below. Here, we show that this bottleneck can be overcome by low dose X-ray speckle visibility spectroscopy. Employing X-ray doses of 22 kGy to 438 kGy and analyzing the sparse speckle pattern of count rates as low as $6.7 \times 10^{-3}$ per pixel we follow the slow nanoscale dynamics of an ionic liquid (IL) at the glass transition. At the pre-peak of nanoscale order in the IL we observe complex dynamics upon approaching the glass transition temperature $T_G$ with a freezing in of the alpha relaxation and a multitude of milli-second local relaxations existing well below $T_G$. We identify this fast relaxation as being responsible for the increasing development of nanoscale order observed in ILs at temperatures below $T_G$.*

**Introduction**

Soft matter, such as for example supercooled liquids, can display relatively slow dynamics on atomic or nanometer length scales. Measuring nanoscale dynamics on time scales of seconds to milliseconds constitutes a considerable experimental challenge. It can be addressed directly by X-ray photon correlation spectroscopy (XPCS) experiments employing coherent X-ray beams [1-7] by tracing fluctuations in X-ray speckle patterns. However, the highly intense X-ray beams of 3rd generation storage rings can also be the cause of considerable radiation damage to the samples. Atomic scale XPCS experiments use X-ray doses of MGy to GGy which can lead to beam induced dynamics even in hard condensed matter samples [8]. Soft and biological matter samples are much more sensitive to radiation damage rendering XPCS experiments with MGy X-ray doses impossible. Overcoming this bottle neck of radiation damage is even more important for the upcoming diffraction limited storage rings (DLSR) which provide an increase in X-ray brilliance of up to two orders of magnitude [2,9,10]. The remedy for beam damage is frequent sample replacement as can be realized using e.g. flow cells [11] or changing the exposure spot on the sample. However, low dose XPCS experiments often show very noisy correlation functions which do not yield conclusive insights into the dynamics or did not extend beyond feasibility studies [12,13].

Here, we demonstrate that X-ray speckle visibility experiments with nanometer resolution can be performed with very low noise levels at radiation doses as low as few kGy. The concept consists in spreading the dose needed to obtain a correlation function over the entire sample volume by measuring at every spot on the sample the visibility of a speckle pattern as a function of exposure time. Mitigating the absorbed dose in this way is done at the expense of signal strength; the collected speckle patterns are sparse with only $10^{-2}$ photons per pixel and possibly even less. We show that the speckle visibility correlation function can nevertheless be extracted by proper assignment of photon probabilities using a sufficiently large number of images.

Ionic liquids (ILs) are molten salts which show a structural heterogeneity on the nanoscale [14,15] . This nanoscale segregation or domain formation is evidenced by a characteristic pre-peak in diffraction experiments at Q-positions around Q=2-3 nm$^{-1}$, much smaller than the molecular structure factor peak. The intensity of this pre-peak increases with decreasing temperature with the intriguing finding that it does not saturate at $T_G$ but instead this nanoscale correlation is found to increase even at temperatures well below $T_G$ [16]. Neutron spin echo (NSE) experiments revealed that this pre-peak is accompanied by complex heterogeneous dynamics with multiple relaxation channels in the picosecond to nanosecond regime at temperatures well above $T_G$ [16]. However, the dynamics at and below the glass transition is occurring on time scales too slow for NSE experiments and has thus so far not been investigated. Applying our scheme, we are able to investigate the nanoscale dynamics of an imidazolium-based ionic liquid (C8mimCl, $C_{12}H_{23}ClN_2$) around its glass transition temperature.

The self-assembled nanoscale order of ionic liquids is based upon a delicate balance between electrostatic and van-der Waals forces and thus very sensitive to radiation damage. Hence, this order quickly decays under the highly intense coherent X-ray beam of 3$^{rd}$ generation storage rings, which requires to optimize the signal to noise ratio (SNR) in such experiments as much as possible. In XPCS an isochronous sequence of N speckle patterns is recorded at the same spot on the sample and used to calculate the intensity autocorrelation function between the images for every pixel according to

$$g_2(i) = \frac{1}{N-i}\sum_{n=1}^{N-i} I(n)I(n+i)/\langle I\rangle^2 = 1 + \beta(Q)|f(Q,\tau)|^2. \text{ Eq. (1)}$$

Here $i$ denotes the temporal separation between two frames of the recorded time series, $f(Q,\tau)$ is the intermediate scattering function (ISF) and $\beta(Q)$ the (Q-dependent) X-ray speckle contrast. The maximum available scattering intensity $k_{max}$ per pixel before the damage threshold is reached is thus distributed onto N images of intensity $k_{max}/N$ each. The SNR in XPCS experiments is [17] (see details in the supplement, Eq. (24))

$$SNR \sim \frac{\beta \, k_{max}}{N} \frac{\sqrt{N_{pairs}N_{pix}N_{rep}}}{\sqrt{2(\beta+1)}} \quad \text{Eq. (2)}$$

where $N_{pairs}$ denotes the number of image pairs available for correlating at this specific time delay, $N_{pix}$ the number of pixels and $N_{rep}$ the number of repeats of this series. Maximizing the SNR requires to choose N as small as possible while still being able to cover the full time interval $(N-1)\cdot\tau$ of interest. We especially note that Eq. (3) implies that an increase in N by a factor of $x$ needs to be counterbalanced by increasing $N_{rep}$ by a factor of $x^2$ which is the decisive factor for the feasibility of such experiments. Assuming typical values of $N_{pix}\sim10^6$, $k_{max}=10^{-2}$, $N=100$ and $\beta=0.1$ we find that the SNR is on the order of one, even for $N_{rep}\sim1000$. Because experiments usually require the systematic change of additional parameters such as temperature or sample composition, increasing $N_{rep}$ by orders of magnitude quickly runs into practical problems with a limited amount of experimental time available.

The SNR can be maximized by utilizing the full available scattering intensity $k_{max}$ for retrieving dynamic information from just a single image by X-ray speckle visibility spectroscopy (XSVS). In this case single images are taken with different exposure times $t_e$ at different sample positions. The XSVS correlation function is then determined by analyzing the contrast of the smeared out speckle patterns as a function of exposure time $t_e$. The speckle contrast is given by [18]

$$\beta(Q,t_e) = \frac{\beta_0}{t_e}\int_0^{t_e} 2\left(1-\frac{t}{t_e}\right)|f(Q,\tau)|^2 d\tau \quad \text{Eq. (3)}$$

which yields access to the temporally averaged ISF. Ref. [18] shows examples for retrieving the ISF from Eq. (3). The SNR of XSVS is [19]

$$SNR = \beta\, k_{max}\sqrt{\frac{N_{pix}N_{rep}}{2(\beta+1)}} \quad \text{Eq. (4)}$$

(supplement Eq. (18)) which can be considerably larger than the one of XPCS, depending on the choice of number of images and delay times. XSVS also allows easily scanning of a non-linear time window as it only depends on the adjustable exposure time. While the temporal averaging of the ISF represents of course a loss of information compared to XPCS, we point out that for low dose experiments XSVS may be the only way to obtain any dynamic information at all. We discuss the details of the respective SNR of both methods in the supplement and note that more intensity per frame is also advantageous to suppress the influence of systematic noise and background effects that are not due to simple Poisson statistics.

**Theory of low intensity X-ray speckle visibility spectroscopy**

The requirements of tracing ms dynamics and working with low radiation doses result in speckle patterns of very low intensities on the order of $10^{-2}$ photons/pixel and below. The speckle contrast from such sparse images can be retrieved by analyzing the photon statistics. The probability of detecting k photons during an exposure time $t_e$ is given by the negative binominal distribution (NBD) [20]

$$P(k,t_e) = \frac{\Gamma(k+M(t_e))}{\Gamma(M(t_e))\Gamma(k+1)}\left(1+\frac{M(t_e)}{\bar{k}}\right)^{-k}\left(1+\frac{\bar{k}}{M(t_e)}\right)^{-M(t_e)}. \quad \text{Eq. (5)}$$

The average photon probability $\bar{k}$ is given by $n_{ph}/n_{pix}$ with $n_{ph}$ denoting the total number of photons registered by the $n_{pix}$ pixels of interest. $M(t_e)$ is the number of modes in a single speckle pattern which is connected to the speckle contrast via $\beta(t_e) = \frac{1}{M(t_e)}$. Its value can be determined for instance from the ratio of zero vs. one photon events $R_{0,1}(t_e)$ (see supplemental material)

$$\beta(t_e) = P(0, t_e) / P(1, t_e) - 1/\bar{k} = R_{0,1}(t_e) - 1/\bar{k} \qquad \text{Eq. (6)}$$

which is the left hand side of Eq. (3).

**Experiment**

The experiment has been performed at the P10 coherence application beamline at PETRA III at DESY. A photon energy of 13 keV has been used with a beam size of 4 x 3 µm² full width at half maximum. The estimated flux of the partially coherent beam is $5 \times 10^{10}$ photons/s. A single 50 ms exposure yielded on average an intensity of $6.7 \times 10^{-3}$ photons/pixel. Temperatures of T=190 - 235 K around the glass transition have been achieved by a liquid nitrogen cryostat. Photons have been detected by the EIGER 4M detector [21] (pixel size of 75 x 75 µm²) of which $8.08 \times 10^5$ pixels have been used for the dynamic analysis. To reduce beam damage, we systematically scanned the sample through the X-ray beam by steps of 10 µm taking a single exposure for each exposure time $t_e$ and sample position. For every temperature, we have measured ten exposure times varying between 50 ms and 1000 ms. For every exposure time 1000 speckle patterns have been measured.

We used the ionic liquid 1-octyl-3-methylimidazolium chloride ($C_{12}H_{23}ClN_2$). The mass absorption coefficient of the IL sample at 13 keV photon energy is 5 cm²/g. The sample was filled into 1 mm diameter quartz glass capillaries and sealed with epoxy resin. Calorimetric studies have shown that the glass transition of the sample is $T_G$ = 214 K [22]. The maximum speckle contrast in our experimental configuration is 0.1.

**Results**

Averaging 1000 diffraction patterns with 1 s exposure each yields the well-known pre-peak in the structure factor of ionic liquids with the maximum of the correlation located at Q=2.6 nm⁻¹ (Fig. 1a). In crystallography the intensities and widths of Bragg peaks are sensitive markers of the loss of order due to beam damage helping to monitor even tiny radiation induced structural changes [23]. The situation is different for disordered samples such as liquids and glasses with no long-range order present, providing only broad diffraction features. In addition, the amount and sensitivity of beam damage depends on the chemical environment. For example, water solutions are very difficult to investigate due to the hydrolysis of water [24]. Nevertheless, often radiation damage manifests itself by more or less abrupt changes to the scattering intensities as for example observed for SAXS experiments from protein solutions [25].

For determining the radiation damage threshold we continuously exposed the sample on a fixed spot and acquired frames with an exposure time of 75 ms at a temperature of T=205 K, i.e. below the glass transition temperature. Using the estimated X-ray flux and calculated absorption properties of the ionic liquid we can convert the exposure time into the corresponding X-ray dose. To calculate the dose we used the estimated flux of $5\times10^{10}$ photons per second on the sample. About 80% of the incoming intensity is absorbed by the wall of the capillary with a wall thickness of 100 µm. For the dose calculation we assume that the total number of absorbed photons spreads over a volume of 7.6x10.2x800 µm³ which corresponds to $3\sigma$ of the Gaussian beam profile. Integrating the scattering signal over the Q-range indicated in Fig. 1a yields the scattering power as a function of absorbed X-ray dose (Fig. 1b). A loss in scattered intensity of the nanoscale structure is detected starting at a dose of 2 MGy. The dose at which the intensity decreases by a factor of two is 7.15 MGy at a real space resolution of $2\pi/Q \sim$ 2.4 nm.

In protein crystallography the maximum tolerable dose depends on the resolution aimed for and is typically around 10 MGy/Å for cryogenically cooled protein crystals [23]. In contrast for our IL the ratio of dose and resolution is more than one order of magnitude smaller with 0.3 MGy/Å pointing to the radiation sensitivity of the IL. In the dynamic experiment, we used doses ranging from 22 kGy to a maximum value of 438 kGy. However, dynamic properties may be affected at doses below the threshold of structural damage [8]. Experiments with comparable dose have shown that the energy spreads over a large volume of the sample and the expected heating is of the order of 10 K [26].

The dynamics of the sample can be inferred from the change of the photon probabilities as a function of exposure time $t_e$. There are two contributions to the change of photon probabilities in X-ray speckle visibility: the first one is the trivial scaling of the average intensity per pixel $\bar{k} = r \cdot t_e$ with $r$ denoting the incoming photon rate. The second factor is the change in mode number $M(t_e)$ which is associated with the dynamics of the sample. In Fig. 2 we plot the difference between the measured zero (blue) and one photon-probabilities (red) and the calculated ones only assuming a simple scaling of the average intensity (T=200 K). Both curves in Fig. 2 deviate from zero and change with $t_e$ indicating that the number of modes $M(t_e)$ is changing during the exposure. Specifically, we note that the number of detected 0-photon events is decreasing with increasing exposure time and the number of 1-photon events is increasing, in agreement with an increasing mode number due to the dynamics of the sample. We observe a similar behavior for the 2- and 3- photon events.

To determine the correlation function, we calculate the ratio of the 0-, and 1-photon events and the averaged photon probability $\bar{k}$ for all speckle images using the scattering signal within the area marked by the red lines in Fig. 1. From these values, we can obtain the speckle contrast $\beta$ using Eq. (6). This contrast value is calculated for the 1000 recorded images for each exposure time $t_e$ and for each temperature T. The error bar is determined from the variance of the 1000 contrast values. Two examples for the contrast $\beta$ calculated from the ratio of 0- and 1-photon events for temperatures of T=220 K and 200 K are shown in Fig. 3. Both contrasts drop as a function of exposure time with the 200 K data providing a much slower loss of speckle visibility indicating slower dynamics at lower temperatures. The correlation function at T=220 K approaches a contrast value of zero at large values of the exposure time indicating that the speckle patterns are fully decorrelated at exposure times beyond 1 s. At lower temperatures, however, the dynamics is partially frozen on the time scales of seconds and beyond as the correlation functions do not approach the value of zero. It is also apparent that there are several relaxation mechanisms involved in the low temperature dynamics of the ionic liquid. At T=200 K for example, in the glassy phase, we observe a fast relaxation with time scales below 100 ms, a second relaxation with time scales of 400 ms and a frozen-in component. We label the three observed processes by P1, P2 and P3 and the solid lines shown in Fig. 3 represents fits using an exponential function and a fixed Kohlrausch-Williams-Watts (KWW) stretching exponent of 0.7 for all three processes [27]. Unfortunately, the limited time window prevents a self-consistent refinement of the three different processes over the whole temperature range. We note, however, that multiple relaxation mechanisms have also been observed in the picosecond to nanosecond regime in the liquid phase at temperatures well above $T_G$ [16].

The complete picture of nanoscale dynamics at the glass transition can be inferred from Fig. 4a showing the speckle contrast over the whole measured temperature interval around $T_G$. The appearance of a frozen-in component is apparent for time scales of 1000 ms and at temperatures of 210 K and below (see also contour plot Fig. 4b). This is where the system falls out of equilibrium and the slowest relaxation gets arrested. This relaxation is attributed to an overall (alpha) type relaxation of the nanoscale order in the IL [28]. However, we still observe dynamics on time scales of 100 - 200 ms in the glassy phase. As this motion does not fully decorrelate the speckle contrast it must be local in nature. However, the fact that it still contributes a rather large component to the decorrelation of the dynamic structure factor at $T_G$ indicates that the associated motion has a considerable spatial extension. The microscopic origin of this mode is not clear but one can relate it to the delicate balance of electrostatic and van-der Waals interactions from the ions and alkyl chains in the IL. This still allows for relaxation mechanisms on the nanoscale which do not freeze out completely below $T_G$. It is tempting to associate the observed increasing nanoscale order below $T_G$ with the presence of this fast localized motion [28].

The speckle contrast at the largest available delay time of 1 s is a measure of the arrested component P3 of the dynamics. The temperature dependence of this component is plotted in Fig. 5(a) showing a sudden increase at a temperature which we associate with the glass transition. The maximum value of the measured frozen-in contrast is 0.017 which, when normalized by the maximum contrast of 0.1, yields a non-ergodicity parameter [29,30] of $f_c = 0.17$ at 200 K (Fig. 5a). Connecting this to the Debye-Waller factor via $f_c = exp(-Q^2\langle u^2\rangle/3)$ we calculate the localization length characterizing the range of motion associated with this fast mode. We find a value of $\sqrt{\langle u^2\rangle} \approx 0.85$ nm which is smaller than the nanostructure correlation length of $L_{cor} \approx \frac{2\pi}{Q} = 2.4$ nm. In fact, this value is closer to typical charge alternation correlation lengths in the IL structure which appear at distances between 0.6 and 0.9 nm. Thus, this fast motion would be in the charge alternation domains and not be related to the apolar domains.

As we cannot disentangle the three processes over the whole temperature range we chose a different way to obtain further insight into the origin of the here observed dynamics. We evaluate the first moment of the non-frozen components of the correlation functions representing a mean decorrelation time for each temperature. This deduced time constants as a function of temperature are shown in Fig. 5b. Close to $T_G$ we can disentangle frozen in and still active motions. The fast averaged motion show an activation energy of (25±5) kJ /mol which is in the range of values reported for different relaxation processes in the liquid phase such as ionic diffusion processes, nanostructural relaxation and local reorientation of the alkyl chains and the imidazolium rings [28].

In conclusion, by spreading the dose over a large sample area and evaluating single photon events we demonstrate that XSVS experiments can be performed at X-ray doses significantly smaller than necessary for typical XPCS experiments. Using this method, we observe the nanoscale dynamics of an ionic liquid below the glass transition. The relaxation in the glassy phase is not completely frozen. Instead we find different processes which are still active in the glassy phase on times scales of 100 ms and below besides a frozen-in alpha type relaxation. This fast dynamic is associated with the peculiar nanostructure of the ionic liquids which still allows for motion below the calorimetric glass transition. It also becomes apparent that for a deeper understanding of the glass transition a larger range of time scales need to be probed which becomes available with the new diffraction limited storage rings (DLSR).

Finally, we note that this approach is an important step forward to measure dynamics in biological samples on nanometer length scales. While the dose used at the shortest exposure of 22 kGy is still higher than the tolerable dose of most proteins, some larger proteins can endure doses of up to 8 kGy [31] which is certainly within the range of the presented approach. With larger scattering intensities at smaller Q-values we see no principle obstacle to perform XSVS experiments with doses of 1 kGy and below.

**Acknowledgements**

This work is supported by the BMBF under project 05K13PS5 and the Swedish Research Council within the Röntgen-Angström-Cluster "Soft matter in motion". Parts of this research were carried out at PETRA III at DESY, a member of the Helmholtz Association.

**References**

[1] J. Carnis *et al.*, Sci Rep **4**, 6017 (2014).
[2] O. Shpyrko, Journal of Synchrotron Radiation **21**, 1057 (2014).
[3] V. M. Giordano and B. Ruta, Nature Communications **7**, 10344 (2016).
[4] B. Ruta, G. Baldi, Y. Chushkin, B. Rufflé, L. Cristofolini, A. Fontana, M. Zanatta, and F. Nazzani, Nature Communications **5** (2014).
[5] B. Ruta, Y. Chushkin, G. Monaco, L. Cipelletti, E. Pineda, P. Bruna, V. M. Giordano, and M. Gonzalez-Silveira, Phys Rev Lett **109**, 165701 (2012).
[6] M. Leitner, B. Sepiol, L.-M. Stadler, B. Pfau, and G. Vogl, Nat Mater **8**, 717 (2009).
[7] G. Grübel and F. Zontone, Journal of Alloys and Compounds **362**, 3 (2004).
[8] B. Ruta, F. Zontone, Y. Chushkin, G. Baldi, G. Pintori, G. Monaco, B. Rufflé, and W. Kob, Scientific Reports **7**, 3962 (2017).
[9] M. Eriksson, J. F. van der Veen, and C. Quitmann, Journal of Synchrotron Radiation **21**, 837 (2014).
[10] E. Weckert, IUCrJ **2**, 230 (2015).
[11] A. Fluerasu *et al.*, Journal of Synchrotron Radiation 15, 378 (2008)
[12] P. Vodnala *et al.*, AIP Conference Proceedings **1741**, 050026 (2016).
[13] C. DeCaro *et al.*, Journal of Synchrotron Radiation **20**, 332 (2013).
[14] A. Triolo, O. Russina, H.-J. Bleif, and E. Di Cola, The Journal of Physical Chemistry B **111**, 4641 (2007).
[15] R. Hayes, G. G. Warr, and R. Atkin, Chem Rev **115**, 6357 (2015).
[16] M. Kofu, M. Nagao, T. Ueki, Y. Kitazawa, Y. Nakamura, S. Sawamura, M. Watanabe, and O. Yamamuro, J Phys Chem B **117**, 2773 (2013).
[17] D. Lumma, L. B. Lurio, S. G. J. Mochrie, and M. Sutton, Review of Scientific Instruments **71**, 3274 (2000).

[18]     R. Bandyopadhyay, A. S. Gittings, S. S. Suh, P. K. Dixon, and D. J. Durian, Review of Scientific Instruments **76**, 93110 (2005).
[19]     S. O. Hruszkewycz *et al.*, Physical Review Letters **109**, 185502 (2012).
[20]     J. W. Goodman, *Statistical optics, p.476* (Wiley, New York, 1985), Wiley series in pure and applied optics.
[21]     I. Johnson *et al.*, Journal of Instrumentation **9**, C05032 (2014).
[22]     O. Yamamuro and M. Kofu, IOP Conference Series: Materials Science and Engineering **196**, 12001 (2017).
[23]     M. R. Howells *et al.*, J Electron Spectros Relat Phenomena **170**, 4 (2009).
[24]     W. M. Garrison, Chemical Reviews **87**, 381 (1987).
[25]     S. Kuwamoto, S. Akiyama, and T. Fujisawa, Journal of Synchrotron Radiation **11**, 462 (2004).
[26]     E.H. Snell *et al.*,  Journal of Synchrotron Radiation **14**, 109 (2007).
[27]     A. Madsen, R. L. Leheny, H. Guo, M. Sprung, and O. Czakkel, New Journal of Physics **12**, 55001 (2010).
[28]     M. Kofu, M. Tyagi, Y. Inamura, K. Miyazaki, and O. Yamamuro, The Journal of Chemical Physics **143**, 234502 (2015).
[29]     L. Berthier and G. Biroli, Reviews of Modern Physics **83**, 587 (2011).
[30]     W. van Megen, S. M. Underwood, and P. N. Pusey, Phys Rev Lett **67**, 1586 (1991).
[31]     C. M. Jeffries *et al.*, Journal of Synchrotron Radiation **22**, 273 (2015).

**Fig.1**

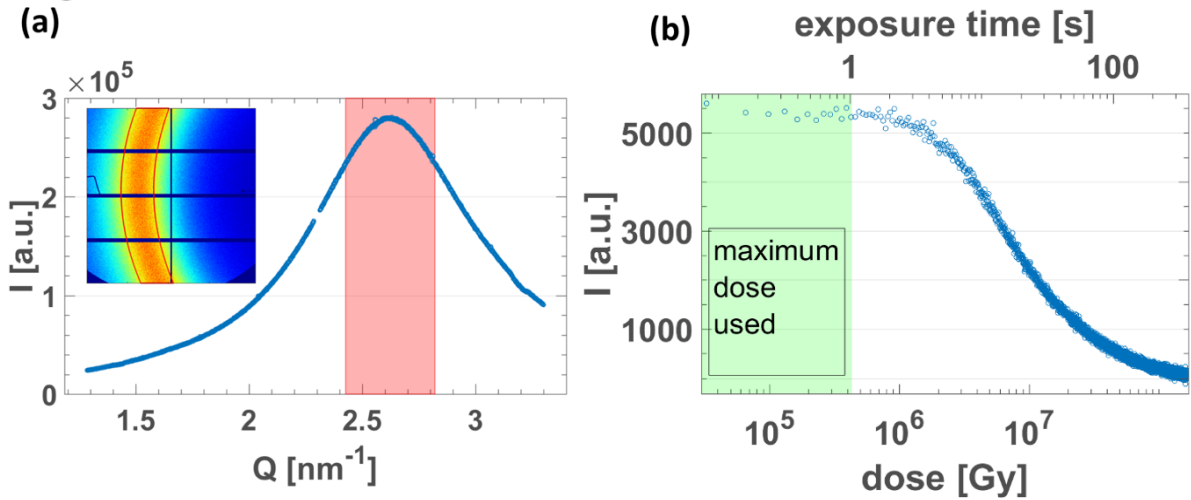

Fig.1:(a) X-ray diffraction pattern from the supercooled ionic liquid C8mimCl associated with the nanoscale order. The inset shows the diffraction pattern as recorded on the area detector. The red box indicates the Q-range used for the XPCS analysis. (b) The integrated X-ray signal from the area of the red box in (a) plotted as a function of the absorbed X-ray dose/exposure time. The green area indicates the range of doses used in the experiment.

**Fig.2**

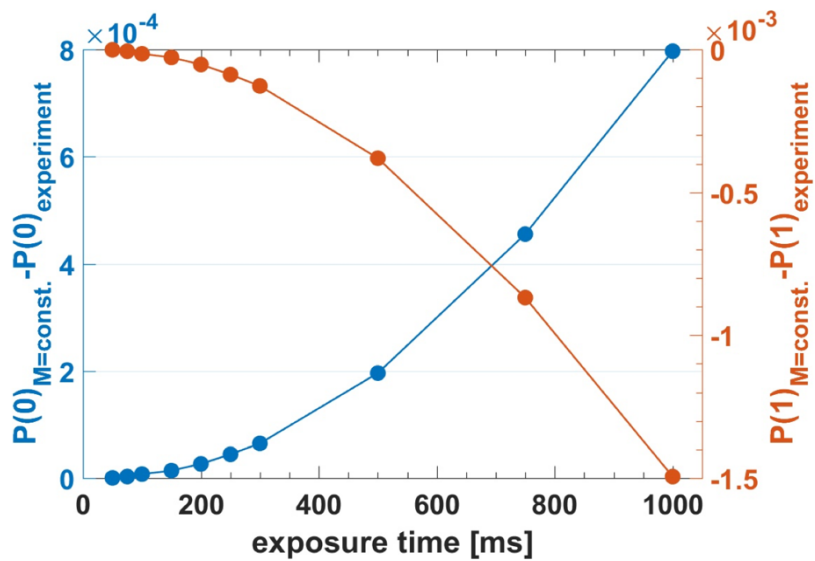

Fig. 2: Blue data: The difference between the actual measured 0-photon probability and the 0-photon probability calculated assuming no dynamics but a simple increase of the mean count rate according to $\bar{k} = r \cdot t_e$. The decreasing number of 0-photon events indicates the decreasing speckle contrast as a function of exposure time caused by the sample dynamics. Red data: same plot as in blue but here for the 1-photon probabilities.

**Fig.3**

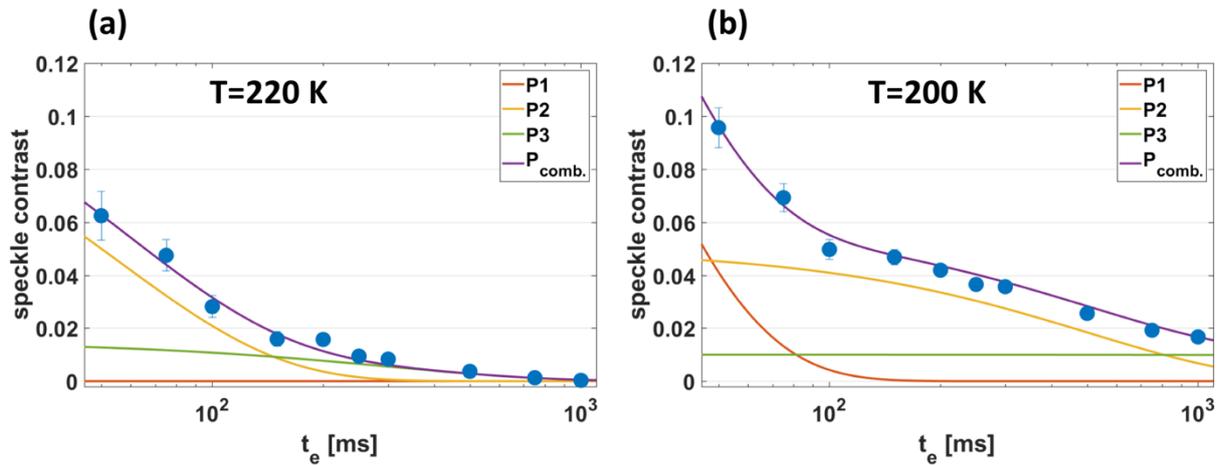

Fig.3: Speckle contrast of the ionic liquid as a function of exposure time for temperatures of 220 K and 200 K and Q=2.6 nm$^{-1}$, respectively. The solid lines are assignments to three different processes labeled P1, P2 and P3. The calorimetric glass transition temperature is 214 K.

**Fig.4**

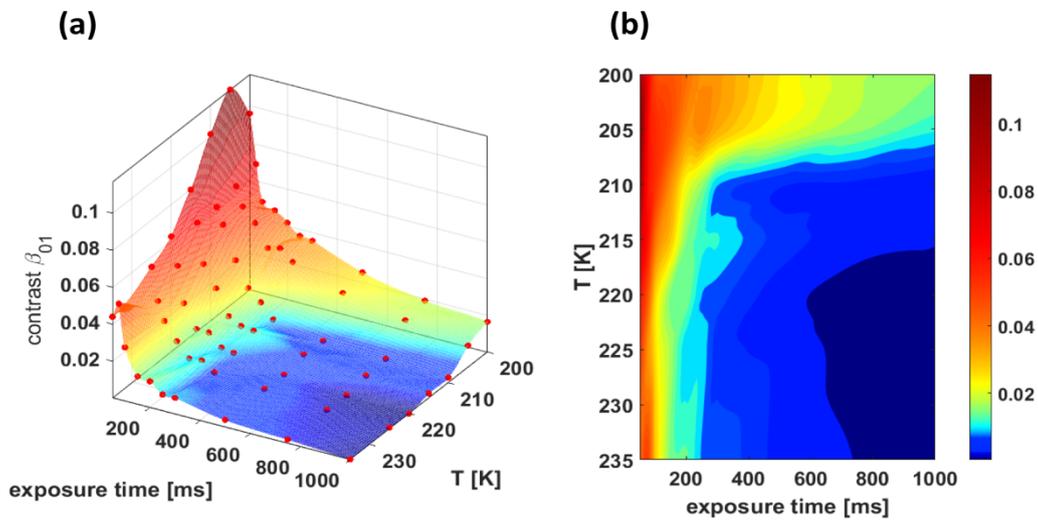

Fig.4: (a) Speckle contrast as a function of temperature and delay time for the ionic liquid C8mimCl at Q=2.6 nm$^{-1}$. The dynamics is progressively slowing down upon approaching $T_G$ with a separation in a fast and slow frozen-in dynamics visible at the lowest temperatures. (b) Contour plot of the pattern shown in (a).

**Fig.5**

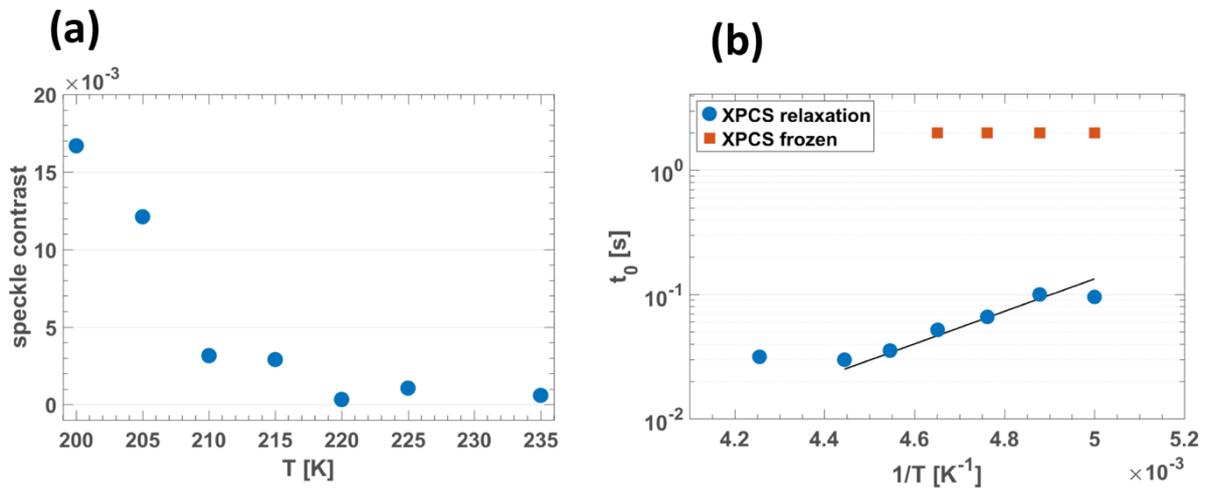

Fig. 5: (a) The X-ray speckle contrast at $t_e = 1$ s exposure time as a function of temperature. (b) Temperature dependence of the averaged decorrelation time of the non-frozen part in the IL (blue dots). The red squares represent the appearance of the frozen in components (i.e. dynamics slower than 1 s). The black line represents an Arrhenius fit with an activation energy of 25 kJ/mol.